\documentclass[twocolumn,superscriptaddress,prb]{revtex4}

\usepackage{graphicx}

\begin{document}

\title{Contact gating at GHz frequency in graphene}

\author{Q. Wilmart}
\affiliation{Laboratoire Pierre Aigrain, Ecole Normale Sup\'erieure-PSL Research University,
CNRS, Universit\'e Pierre et Marie Curie-Sorbonne Universit\'es,
Universit\'e Paris Diderot-Sorbonne Paris Cit\'e, 24 rue Lhomond, 75231 Paris Cedex 05, France}

\author{A. Inhofer}
\affiliation{Laboratoire Pierre Aigrain, Ecole Normale Sup\'erieure-PSL Research University,
CNRS, Universit\'e Pierre et Marie Curie-Sorbonne Universit\'es,
Universit\'e Paris Diderot-Sorbonne Paris Cit\'e, 24 rue Lhomond, 75231 Paris Cedex 05, France}

\author{M. Boukhicha}
\affiliation{Laboratoire Pierre Aigrain, Ecole Normale Sup\'erieure-PSL Research University,
CNRS, Universit\'e Pierre et Marie Curie-Sorbonne Universit\'es,
Universit\'e Paris Diderot-Sorbonne Paris Cit\'e, 24 rue Lhomond, 75231 Paris Cedex 05, France}

\author{W. Yang}
\affiliation{Laboratoire Pierre Aigrain, Ecole Normale Sup\'erieure-PSL Research University,
CNRS, Universit\'e Pierre et Marie Curie-Sorbonne Universit\'es,
Universit\'e Paris Diderot-Sorbonne Paris Cit\'e, 24 rue Lhomond, 75231 Paris Cedex 05, France}

\author{M. Rosticher}
\affiliation{D\'epartement de Physique, Ecole Normale Sup\'erieure-PSL Research University,
CNRS, Universit\'e Pierre et Marie Curie-Sorbonne Universit\'es,
Universit\'e Paris Diderot-Sorbonne Paris Cit\'e, 24 rue Lhomond, 75231 Paris Cedex 05, France}

\author{P. Morfin}
\affiliation{Laboratoire Pierre Aigrain, Ecole Normale Sup\'erieure-PSL Research University,
CNRS, Universit\'e Pierre et Marie Curie-Sorbonne Universit\'es,
Universit\'e Paris Diderot-Sorbonne Paris Cit\'e, 24 rue Lhomond, 75231 Paris Cedex 05, France}

\author{N. Garroum}
\affiliation{Laboratoire de Physique Statistique, Ecole Normale Sup\'erieure-PSL Research University,
CNRS, Universit\'e Pierre et Marie Curie-Sorbonne Universit\'es,
Universit\'e Paris Diderot-Sorbonne Paris Cit\'e, 24 rue Lhomond, 75231 Paris Cedex 05, France}

\author{G. F\`eve}
\affiliation{Laboratoire Pierre Aigrain, Ecole Normale Sup\'erieure-PSL Research University,
CNRS, Universit\'e Pierre et Marie Curie-Sorbonne Universit\'es,
Universit\'e Paris Diderot-Sorbonne Paris Cit\'e, 24 rue Lhomond, 75231 Paris Cedex 05, France}

\author{J-M. Berroir}
\affiliation{Laboratoire Pierre Aigrain, Ecole Normale Sup\'erieure-PSL Research University,
CNRS, Universit\'e Pierre et Marie Curie-Sorbonne Universit\'es,
Universit\'e Paris Diderot-Sorbonne Paris Cit\'e, 24 rue Lhomond, 75231 Paris Cedex 05, France}

\author{B. Pla\c{c}ais}
\email{bernard.placais@lpa.ens.fr}
\affiliation{Laboratoire Pierre Aigrain, Ecole Normale Sup\'erieure-PSL Research University,
CNRS, Universit\'e Pierre et Marie Curie-Sorbonne Universit\'es,
Universit\'e Paris Diderot-Sorbonne Paris Cit\'e, 24 rue Lhomond, 75231 Paris Cedex 05, France}

\begin{abstract}

The paradigm of graphene transistors is based on the gate modulation of the channel carrier density by means of a local channel gate. This standard architecture is subject to the scaling limit of the channel length and further restrictions due to access and contact resistances impeding the device performance. We propose a novel design, overcoming these issues by implementing additional local gates underneath the contact region which allow a full control of the Klein barrier taking place at the contact edge. In particular, our work demonstrates the GHz operation of transistors driven by independent contact gates. We benchmark the standard channel and novel contact gating and report for the later dynamical transconductance levels at the state of the art. Our finding may find applications in electronics and optoelectronics whenever there is need to control independently the Fermi level and the electrostatic potential of electronic sources or to get rid of cumbersome local channel gates.

\end{abstract}

\maketitle


 In general, contacts behave as passive elements impeding the performance of electronic devices. In high mobility graphene transistors the contact resistance competes with the channel resistance already for channel lengths in the hundred nanometer range \cite{Wu2012nl}. A standard route to minimize this spurious contribution is to match contact doping with  channel doping using an overall back gate. Introduced in silicon and carbon nanotube Schottky barrier transistors \cite{Sze2007wiley,Heinze2002prl}, back gating proves particularly efficient in graphene \cite{Xia2011nnano,Berdebes2011ieee,Knoch2012ieee} and other 2D materials \cite{Liu2015acs}, thanks to the weak screening at 2D. Going one step further, one can think of turning contacts to active elements by controlling independently and dynamically their electric and chemical potentials using individual contact gates; the non linear element is then the Klein tunneling barrier that develops at the contact edge due to the work function mismatch between metal and graphene \cite{Giovanetti2008prl}. Such a contact gating is challenging as it requires a set of local gates to engineer the doping profile all along the graphene sheet with the Fermi wave length resolution. In the present work we have realized dual gate Klein barrier transistors (KBTs) achieving these conditions using a nano-patterned bottom gate array, thin hexagonal boron nitride (hBN) dielectric, and palladium (Pd) contacts. The contact gate controls the doping of contacted graphene locally, including its polarity, which allows operating the KBT in a fully tunable bipolar regime. Its low bias DC properties are accurately mapped using a ballistic Klein tunneling junction model \cite{Cayssol2009prb} from which we extract the relevant contact parameters. Its dynamical response is investigated at GHz frequency, where contact gating is benchmarked against conventional channel gating. The realization of gated contacts opens the way to new types of devices such as (channel) gate-free transistors for microwave or photo-detectors \cite{Xia2009nnano,Yu2009nnano}, tunable nanosecond electron sources for quantum electronics \cite{Bocquillon2013science,Dubois2013nature} or spintronics \cite{Seneor2012mrs}, as well as new architectures based on Dirac Fermion optics \cite{Katnelson2006nphys,Lee2015nphys,Wilmart2014_2dm}.

\begin{figure}[hh]
\centerline{\includegraphics[width=8cm]{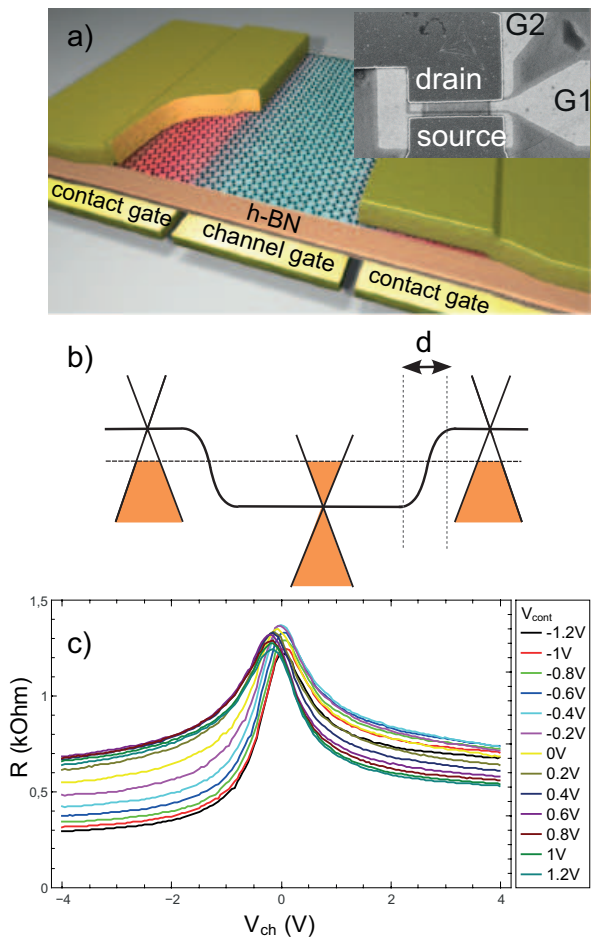}}
\caption{Panel a): artist view of the contact-gated graphene transistor achieving independent channel and contact gating. Inset : SEM image of the transistor; here G1 and G2 stand for channel and contact gates respectively. The scale is given by the channel width and length of $1.1\;\mathrm{\mu m}$ and $0.2\;\mathrm{\mu m}$ respectively. Panel b) : sketch of the doping profile of the device. Panel c) : Dependence of the resistance transfer function by the contact-gate voltage. Characteristic features of independent contact and channel gating are prominent : \emph{i.e.} a strong modulation in the p and n doped regimes and the absence of modulation of the Dirac peak position.}\label{Wilmart.fig1}
\end{figure}

 A graphene contact is a composite element made of two junctions in series: a vertical metal-graphene junction and an in-plane contact junction. The former stems from the momentum mismatch between zone-center metallic and zone-edge graphene electrons and the latter results from doping gradients at the contact edge. According to Giovanetti et al. \cite{Giovanetti2008prl}, a charge transfer takes place at the metal-graphene junction as a result of work function imbalance $\Delta W$  between graphene and metal; its equilibrium density is governed by $\Delta W$ and the screening energy $\epsilon_c$ of graphene (see Eq.(\ref{rouge}) below). However the electrostatic problem is not fully closed in general as the doping may be further controlled by remote electrostatic influence from the back side of the contact. Therefore Eq.(\ref{rouge}) incorporates contact back gating by the voltage $V_{cont}$, which fully sets the Fermi energy of the graphene area located beneath the metal. In-plane junctions have been investigated theoretically (see e.g. Ref.\cite{Cayssol2009prb}) and experimentally \cite{Xia2011nnano,Knoch2012ieee,Wu2012nl}. As a difference with gate defined channel junctions \cite{Huard2007prl,Standler2009PRL,Rickhaus2013ncomm}, contact junctions are generally sharp and  ballistic; their resistance is determined by the junction length $d$ and the doping (polarity and concentration) on both sides. With $k_Fd/2\pi\sim 1$, where $k_F$ is the electronic wave number in the least doped side, sharp p-n junctions have a high transparency $T\gtrsim0.5$ \cite{Cayssol2009prb}.  In this work we demonstrate a Klein barrier transistor, equipped with a bottom channel gate (bias $V_{ch}$) and two interconnected source and drain contact gates  (bias $V_{cont}$) as sketched in Fig.\ref{Wilmart.fig1}-a. The independent dual gate control of Klein barriers is highlighted in the low-bias resistance curve $R(V_{ch})$, where $R$ is strongly modulated by $V_{cont}$ at large channel p- and n-doping (Fig.\ref{Wilmart.fig1}-b) while keeping independent of $V_{cont}$ at channel neutrality. At finite bias, the dual-gate KBT shows a large RF transconductance ($g_{m}^{RF}$ per unit width $W$) upon channel ($g_{m,ch}^{RF}$) and contact  ($g_{m,cont}^{RF}$) gating, both values approaching the state of the art for RF field effect transistors ($g_{m}/V_{ds}\sim 250\;\mathrm{S.m^{-1}V^{-1}}$).

\section{Methods}

  Following Ref.\cite{Meric2011ieee} we use h-BN flakes as a mobility preserving back-gate dielectric (dielectric constant $\kappa\simeq4.2$). A $16\;\mathrm{nm}$-thin h-BN flake is deposited on a $20\;\mathrm{nm}$-thick tungsten film that is nanostructured  with $30\;\mathrm{nm}$ gaps  to realize a back gate array as shown in Fig.\ref{Wilmart.fig1}-a (inset). The 30 nanometer long intergate trenches are achieved by e-beam lithography and dry etching.  Tungsten was used here as a refractory metal preserving the gate pattern resolution during annealing processes. In addition, the h-BN dielectrics shows superior dynamical properties, as compared with  conventional thin oxides,  due to the absence of spurious charge traps. The exfoliated graphene and h-BN flakes are transferred with a dry technique \cite{Gomez2014_2dm}. Palladium contacts are deposited and aligned with the trenches to set independent channel and contact gating (inset of Fig.\ref{Wilmart.fig1}-a). The channel length and width are $L=0.2\;\mathrm{\mu m}$ and $W=1.1\;\mathrm{\mu m}$. Other samples have been fabricated using CVD graphene showing similar behaviors, with however a lower contrast due to longer and more diffusive channels length ($L=0.5\;\mathrm{\mu m}$); for simplicity we restrict ourself on the data obtained on the exfoliated sample. The nanoscale packing of three electrodes (a metallic contact and two gates) provides a vectorial control of the local electric field at the contact edge (Fig.\ref{Wilmart.fig3}-b):  the two normal components, $E_\perp\lesssim 0.5\;\mathrm{V/nm}$ limited by the h-BN dielectric strength  \cite{Hattori2015acsnano}, controlling the doping density on both sides of the junctions, whereas the in-plane component, $E_\parallel\sim\pm 0.2\;\mathrm{V/nm}$ limited by the inter-gate spacing, controls the junction steepness. In our working conditions we have $k_F^{cont}d/2\pi\lesssim 1$. The transistor is embedded in a three port coplanar wave guide (CPW) (Fig.\ref{Wilmart.fig4}-a) used for DC and RF characterization; the transfer curves and scattering parameters are measured identically for contact and/or channel gating. Experiments are carried out in a $40\;\mathrm{GHz}$ variable temperature probe station, with a measuring window limited to a 3 GHz by the finite resistivity ($\sim300\;\Omega$ square) of the thin tungsten metallizations. The experimental data below refer to room temperature measurements, but we have checked that device properties are preserved, and eventually enriched by Fabry-P\'erot oscillations, at cryogenic temperatures \cite{Gorbachev2008nl,Young2009nphys,Wu2012nl}.

\begin{figure}[hh]
\centerline{\includegraphics[width=9cm]{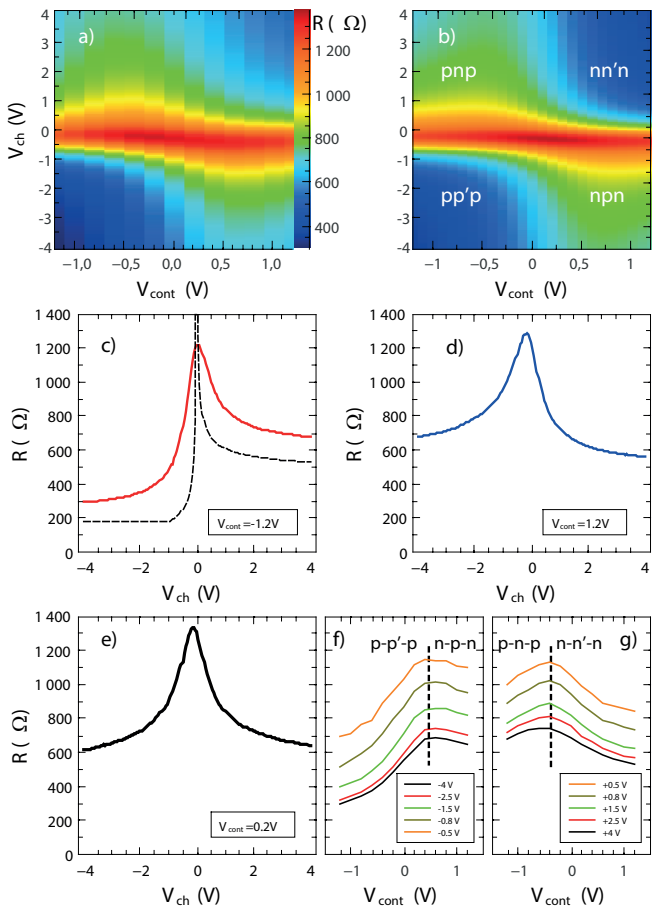}}
\caption{DC operation of the contact gated transistor. Panel a) : color map of the device resistance as a function of contact and channel gate voltages. Panel b) : Corresponding simulation using the fitting model and parameters described in the text. Panels
c-e) : Typical resistance transfer curves for a subset of contact gate voltages. The asymmetry of curves in panels c) and d) reflects the p and n doping of the contact; the symmetry of  curve in panel e) suggests a neutral contact.
The dotted line in panel c) represents the transfer curve expected from a fully ballistic device. Panels f) and g) : Device resistance as a function of contact gate voltage for channel gate voltages  $V_{ch}=-4V\rightarrow-0.5V$ and $V_{ch}=+4V\rightarrow +0.5V$  (from bottom to top) respectively.
 }\label{Wilmart.fig2}
\end{figure}

\section{DC results}

Fig.\ref{Wilmart.fig1}-c shows the resistance transfer curve $R(V_{ch})$ for contact gate voltages $V_{cont}=-1.2\rightarrow1.2\;\mathrm{V}$. As seen in the figure, contact gating efficiently modulates the device resistance away from charge neutrality point (CNP),   $R(V_{ch})$ being modulated by a factor $2.3$ for $V_{ch}<0$ and  $1.3$ for $V_{ch}>0$. By contrast, the CNP conductance remains unchanged indicating a mobility limited conductance ($\mu\simeq6000\;\mathrm{cm^2V^{-1}s^{-1}}$ as explained below). Note that, in contrast with previous investigations using a combination of back and top gates \cite{Knoch2012ieee}, the CNP-peak position is not shifted by the contact gate. The independence of the channel and contact gating not only facilitates the control of contact junctions at DC but also paves the way for new RF operation. For a quantitative analysis of the combined effects of channel and contact gating we consider below the full set of resistance transfer curves.

 A color plot representation of the  $R(V_{ch},V_{cont})$  is presented in Fig.\ref{Wilmart.fig2}-a. As seen in the figure, the experimental data map the four polarities of channel and contact doping. Figs.\ref{Wilmart.fig2}-c,d,e show vertical cuts $R(V_{ch})$ for three representative values of the contact gate voltage: at negative $V_{cont}=-1.2\;\mathrm{V}$ (panel-c) the curve is most asymmetric, indicating a strengthening of the pristine p-doping of Pd contacts. The asymmetry is opposite, although less pronounced, for $V_{cont}=+1.2\;\mathrm{V}$ (panel-d) suggesting the reversal from p- to n-doping of the contacts. At $V_{cont}=0.2\;\mathrm{V}$ (panel-e) the curve is symmetric and corresponds to neutral contacts. The  resistance minimum, $R(-4V,-1.2V)=300\;\mathrm{\Omega}$, is achieved for p-type  contacts and channel doping; it deviates from the ballistic case (dotted line in Fig.\ref{Wilmart.fig2}-c and calculations below) by a small amount, $\simeq110\;\mathrm{\Omega}$, attributed to the metal-graphene resistance ($R_{mg}\simeq55\;\mathrm{Ohms}$ per contact). Our overall contact resistance compares favorably with the best achievements of $100\mathrm{\Omega.\mu m}$ \cite{Xia2011nnano,Zhong2015nr}.  The asymmetry between p-doped and n-doped regimes,  $\Delta R(V_{ch}=\pm4 V, V_{cont}=-1.2 V)\simeq400\;\mathrm{Ohms}$, is itself consistent with the ballistic junction model below. Figs.\ref{Wilmart.fig2}-f,g are $R(V_{cont})$ cuts for  p- and n- channel doping for contact junction dominated regimes. The curves show a maximum and an asymmetry which are typical of ordinary transfer curves, confirming our ability to drive the contact doping through neutrality. These curves are more direct evidence of contact doping reversal than the conventional ones obtained with remote back gates where polarity reversal is signaled by a smeared second peak  \cite{Knoch2012ieee,Zhong2015nr}. The positions of the maxima are independent of $R(V_{ch})$ (given a channel polarity) confirming the small electrostatic coupling between contact and channel gates, the $R(V_{cont})$ curves being mainly shifted upward on decreasing channel doping due the finite carrier mobility contribution of the channel. The main features of the $R(V_{ch},V_{cont})$  data are well reproduced in the plot of Fig.\ref{Wilmart.fig2}-b, obtained using the simple model detailed below.

\begin{figure}[hh]
\centerline{\includegraphics[width=8cm]{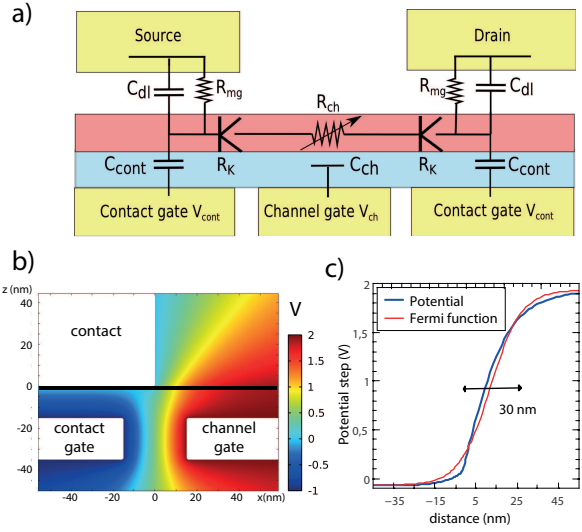}}
\caption{Panel a) Equivalent circuit diagram of the sample. The contact area is represented by a resistance $R_{mg}$ in parallel with an electrical double layer capacitance $C_{dl}$ that stand for electrochemical balance. The doping of graphene underneath the contact is modulated through the capacitance $C_{cont}$. The "K" symbol represents the Klein tunneling junction between contact area and channel area with its associated resistance $R_K(V_{ch},V_{cont})$. $R_{ch}=\frac{L}{Wne\mu}$ is the diffusive resistance of the channel modulated by the channel gate through $C_{ch}$.
Panel b) Electrical potential map of the device at the contact junction from finite element simulation for 2 typical channel and contact gate voltages, $V_{cont}=-1\;\mathrm{V}$ and $V_{ch}=+2\;\mathrm{V}$, the contact being at $V_{source}=0\;\mathrm{V}$. Dimensions are labeled in nanometers, the graphene layer being located at $z=0$. Panel c) Horizontal cut of the potential through the junction together with a Fermi like function. The $10-90\%$ extension is $d\simeq 30nm$.}\label{Wilmart.fig3}
\end{figure}

\section{Model}

To describe the contact properties we use an electrostatic model for the contact doping from Ref.\cite{Giovanetti2008prl} and a Fermi-function potential model for the contact junction resistance from  Ref.\cite{Cayssol2009prb}. The resistance is calculated using the lumped electrical elements description in Fig.\ref{Wilmart.fig3}-a, where the contact junction is represented by a "K" symbol (by reference to Klein tunneling junctions). In the electrostatic model the metal-graphene contact is described by a tunnel resistance $R_{mg}$ in parallel with a metal-graphene capacitance $C_{dl}$ (per unit area), the later accounting for the charge double layer formed at the metal-graphene interface as a result of electrochemical potential mismatch $\Delta W$. The contact gate tunes the double layer charge with a coupling controlled by the contact-gate capacitance  $C_{cont}$. Taking into account quantum capacitance effects \cite{Pallecchi2011prb} (not represented in Fig.\ref{Wilmart.fig3}-a) one obtains the following expression for the chemical potential $E_{F}^{cont}$ of contacted graphene \cite{Giovanetti2008prl}:
\begin{eqnarray}
E_{F}^{cont}&=&\mathrm{sgn}(\epsilon_{W})\times \epsilon_c (1 - \sqrt{1+ 2|\epsilon_w|/\epsilon_c})\\
\epsilon_{W} &=&\Delta W + eV_{cont}\frac{C_{cont}}{C_{cont}+C_{dl}} \nonumber\\
\epsilon_c &=&\frac{\pi\hbar ^2 v_f^2}{2e^2}(C_{dl}+C_{cont}) . \nonumber
\label{rouge}
\end{eqnarray}
The doping polarity is determined by the sign of the effective electrochemical potential mismatch $\epsilon_{W}$, which is tuned by the local back gate voltage $V_{cont}$ according to the capacitance level arm ratio $C_{cont}/C_{dl}$. Ab-initio calculations \cite{Giovanetti2008prl} predict small double layer thickness ($\sim0.3$--$0.5\;\mathrm{nm}$) and accordingly large $C_{dl}\sim15$--$30\;\mathrm{fF/\mu m^2}$. However the metal graphene coupling is process dependent and experiments show large scatter in the contact resistance presumably due to doping variations. In addition our measurements show that, for a given sample, the contact coupling eventually depends also on thermal cycling.  The term $\epsilon_c$ defines the screening energy scale of the gated contact assembly; at low doping ($\epsilon_{W}\ll \epsilon_{c}$), the screening by graphene is weak and the contact gate controls the electrostatic potential of graphene while the gate charge is taken over by the contact itself.  At large doping ($\epsilon_{W}\gg \epsilon_{c}$) the charge is carried by the graphene itself that screens the contacting metal. Eq.(\ref{rouge}) accounts for the crossover between the two regimes. In the channel, the doping is determined by $V_{ch}$ according to the usual parallel plate capacitor model.

Several models have been proposed to calculate the transmission $T(\theta)$ of a potential step as function of the incident angle $\theta$ of carriers to the junction normal \cite{Cheianov2006prb,Sonin2008prb,Cayssol2009prb}. We assume ballistic junctions and rely on the Fermi-function model of Ref.\cite{Cayssol2009prb} where the potential step is described the function $V(x)\propto(1+e^{-x/w})^{-1}$ where $w$ is the step length scale. As a benefit it provides analytic expressions for $T(\theta)$ for arbitrary junction length (see Eq.(10) in Ref.\cite{Cayssol2009prb}).  Using numerical simulation tools (Fig.\ref{Wilmart.fig3}-b), we have checked that the potential step at a contact edge in Fig.\ref{Wilmart.fig3}-c does mimic a Fermi function step with a $10-90 \%$ extension $4w\simeq d\simeq 30\;\mathrm{nm}$. In this simulation we have neglected graphene screening, an assumption that is  valid at low carrier density, and strictly relevant at p-n contact junctions. The resistance $R_K$ in Fig.\ref{Wilmart.fig3}-a is then calculated using  $R_K^{-1}(V_{cont},V_{ch})=\frac{4e^2}{h}\frac{W k_F^{cont}}{\pi}\langle T\rangle_\theta$, where $\langle T\rangle_\theta(V_{cont},V_{ch})$ is the angular average of the junction transmission appropriate for diffusive leads, and $k_F^{cont}$ the Fermi momentum of contacted graphene \cite{Cayssol2009prb}. Restricting ourself to linear conductance and assuming drain-source symmetry and incoherent transport in the channel, the drain and source junction resistances simply add up. Finally we add a channel resistance $R_{ch}$ in Fig.\ref{Wilmart.fig3}-a, which is calculated taking an energy independent mobility $\mu=Const.$. Based on this simple model we can reproduce in Fig.\ref{Wilmart.fig2}-b the main features of the experimental plot $R(V_{ch},V_{cont})$ in  Fig.\ref{Wilmart.fig2}-a. In the simulation we have imposed $C_{ch}=C_{cont}=2.3\;\mathrm{fF/\mu m^2}$ and the junction length $d=30\;\mathrm{nm}$ from device geometry. The fitting parameters are $R_{mg}=55\;\mathrm{Ohm}$, $\mu=6000\;\mathrm{cm^2V^{-1}s^{-1}}$, $C_{dl}=4.5\;\mathrm{fF/\mu m^2}$ ($\epsilon_c=58\;\mathrm{meV}$) and $\Delta W=50\;\mathrm{meV}$. From these numbers we deduce the investigated contact doping range  $E_{F}^{cont}=-135\rightarrow  117\;\mathrm{meV}$. Note that the value of the mobility is consistent with a ballistic length  $l_B=\mu E_F/ev_F\sim60\mathrm{nm}$ (at $E_F\sim0.1\;\mathrm{eV}$) that is larger than the junction length. In addition we have accounted for a gaussian broadening of the density corresponding to an experimental minimum sheet carrier concentration of $n_0\simeq2.5\times10^{11}\;\mathrm{cm^{-2}}$. The above fitting process involves a priori five adjustable parameters, $\Delta W$, $C_{dl}$, $R_{mg}$, $\mu$ and $n_0$. The last two ones are determined as usual by the $R(V_{ch})$ dependence at DP while the three others control $R(V_{ch})$ away from DP. They can be determined thanks to the additional $V_{cont}$ dependence controlling the asymmetry of the $R(V_{ch})$ curves from which we deduce the contact parameters $\Delta W$, $C_{dl}$. As mentioned above, we finally deduce  $R_{mg}$ from the (constant) symmetric part of the $R(V_{ch})$ curves.
A superiority of our dual gate transistor is that it provides direct estimates of the metal-graphene contact parameters, i.e. the work function mismatch and the screening energy.

\begin{figure}[hh]
\centerline{\includegraphics[width=10cm]{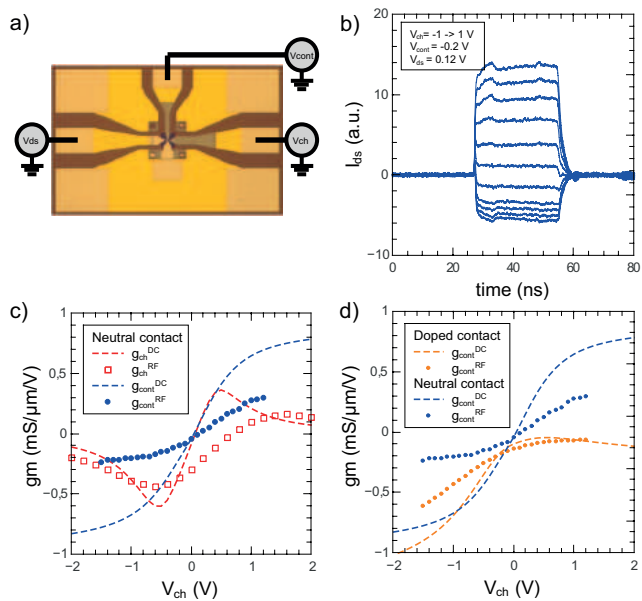}}
\caption{High frequency operation of the contact gated transistor.  Panel a) : Optical image of the 3-port coplanar wave guide (CPW) used in the experiment. CPW dimensions are $600\times400\;\mathrm{\mu m}$; the transistor (inset of Fig.\ref{Wilmart.fig1}-a) is embedded in the central part of the CPW (not visible in the figure). Panel b) : drain current temporal trace under contact-gate pulses of amplitude $\lesssim90\;\mathrm{mV}$ for $V_{ds}=120\;\mathrm{mV}$, $V_{ch}=-1\rightarrow 1\;\mathrm{V}$, $V_{cont}=-0.2\;\mathrm{V}$. Panel c) :  benchmarking of channel and contact gating as measured by the DC and RF transconductances (per unit width and DC drain-source bias voltage). The comparison is carried out in neutral contact conditions $V_{cont}=0\;\mathrm{V}$. Panel d) : effect of contact doping at $V_{cont}=-0.9\;\mathrm{V}$ (orange dots) on the contact gate transconductance by comparison to the neutral case $V_{cont}=0$ (blue dots).}\label{Wilmart.fig4}
\end{figure}

\section{RF results}

Based on the understanding of the DC properties of our dual gate KBT, we proceed by investigating its dynamical properties. Especially, we show here that the performance of the transistor, driven from its contact gate, can be as good as when it is driven from the usual channel gate. Here we are restricted to the $0.1$--$3\;\mathrm{GHz}$ frequency range, due to the combined effects of the gate resistance ($\simeq 1.5\;\mathrm{k\Omega}$) and stray capacitance ($\simeq 50\;\mathrm{fF}$). In this window the transconductance data $g_m$ in Fig.\ref{Wilmart.fig4}-b,c are frequency independent. We have selected  representative gate biasing conditions for both channel- and contact-gating excitations. The sample is biased with a drain-source voltage $V_{ds} < 0.3\;\mathrm{V} $ to achieve a finite $g_m$ while preserving linear conditions $g_m \propto V_{ds}$. At higher biases, velocity saturation effects arise which are beyond the scope of the present work. In Fig.\ref{Wilmart.fig4}-c we compare the RF transconductance for a channel gate excitation $g_{ch}^{RF}$ (red squares) with that of a contact gate excitation $g_{cont}^{RF}$ (blue dots). This benchmarking is performed in the case of neutral contacts ($V_{cont}=0\;\mathrm{V} $) where the device is most symmetric. We first note that channel and contact gating are equally efficient with a transconductance level $g_m /V_{ds}/W \simeq 0.25 mS/\mu m/V$ that matches the current state of the art \cite{Lin2010Science,Meric2011ieee,Pallecchi2011apl}. For a more quantitative account of the dynamics of contact gating we have also plotted in Fig.\ref{Wilmart.fig4}-c the DC transconductances $g_{ch}^{DC}$ and $g_{cont}^{DC}$ (dashed lines). The drop from DC to RF (Fig.4-c,d) is more pronounced for contact gating, presumably due to a larger gate source capacitance. In Fig.\ref{Wilmart.fig4}-d we analyze the influence of contact doping on the contact-gate transconductance; we observe that  $g_{cont}^{RF}$ is strongly enhanced when contacted graphene is driven deep in the hole doping range ($V_{cont}=-0.9\;\mathrm{V} $) reaching a value as large as $0.6 mS/\mu m/V$. These transconductance levels are very promising for applications in RF transistors that would however require low-resistance contact gates.

Finally, we show in Fig.4-b the response of the dual gate KBT under pulsed contact gating. As seen in the figure the sign of output current is reversed upon changing the channel doping polarity (at fixed contact doping) in agreement with the transconductance measurements for similar bias conditions (panel c). A technical but important feature is the fact that the output current pulse is a faithful replica of the contact gate pulse. We have realized similar devices where hBN is replaced by a thin aluminium oxide layer, where the situation is very different with the drain current drooping at the nanosecond
scale after a gate voltage step. We attribute this affect to the relaxation of spurious charges at the graphene-oxide interface. Our hBN devices are devoid of trapped charges and therefore highly suitable for pulsed electronic applications.

\section{Conclusions}

In conclusion we have demonstrated contact gating both at DC and RF, in a dual gate Klein barrier transistor using graphene on boron nitride with a set of local bottom gates. Our device operates in the fully bipolar regime with transport properties that are accurately mapped with a ballistic Klein tunneling junction model. Active contact gating is demonstrated at RF frequency, and compares favorably with conventional channel gating. This finding opens new routes for graphene electronics and optoelectronics by using low resistance bottom gates and/or separate contact gates to tune independently the drain and source doping.

\begin{acknowledgments} The research leading to these results have received partial funding from the European union Seventh Framework programme under grant N:604391 Graphene Flagship, and from the ANR-14-CE08-018-05 "GoBN". The work of Q.W. was supported by a DGA-MRIS scholarship. We thank thank L. Becerra for providing the tungsten films, P.J. Hakonen and G.S. Paraoanu for fruitful discussions.
\end{acknowledgments}


\begin{thebibliography}{}

\bibitem{Wu2012nl}
Y. Wu, V. Perebeinos,Y-M. Lin, T. Low, F. Xia, and P. Avouris, \emph{Nano Lett.} \textbf{12}, 1417 (2012)

\bibitem{Sze2007wiley}
S.M. Sze, K.NG Kwok,
 \emph{Physics of Semicondcutors devices, Wiley ed. } \textbf{Thrird edition}, P.181 (2007).

\bibitem{Heinze2002prl}
S. Heinze, J. Tersoff, R. Martel, V. Derycke, J. Appenzeller,  Ph. Avouris,
 \emph{Phys. Rev. Lett.} \textbf{89}, 106801 (2002).

\bibitem{Xia2011nnano}
F. Xia, V. Perebeinos, Y.-M. Lin, Y. Wu ,  P. Avouris, \emph{Nature Nanotechnology} \textbf{6}, 179 (2011).

\bibitem{Berdebes2011ieee}
D. Berdebes, T. Low, F. Xia, and J. Appenzeller,
\emph{IEEE Trans. Electron Devices} \textbf{58}, 3925 (2011)

\bibitem{Knoch2012ieee}
J. Knoch, Z. Chen, J. Appenzeller,
\emph{IEEE Trans. on Nanotechnology} \textbf{11}, 513 (2012)

\bibitem{Liu2015acs}
W. Liu, D. Sarkar, J. Kang, W. Cao, K. Banerjee,
\emph{ACS Nano} \textbf{9}, 7904 (2015).

\bibitem{Giovanetti2008prl}
G. Giovannetti, P. A. Khomyakov, G. Brocks, V. M. Karpan, J. van den Brink, P. J. Kelly, \emph{Phys. Rev. Lett.} \textbf{101}, 026803 (2008).

\bibitem{Cayssol2009prb}
J. Cayssol, B. Huard, D. Goldhaber-Gordon,  \emph{Phys. Rev. B} \textbf{79}, 075428 (2009),

\bibitem{Xia2009nnano}
F. Xia,T. Mueller, Y-M.  Lin, A. Valdes-Garcia, P. Avouris,
\emph{Nature Nanotech.} \textbf{4}, 839  (2009).

\bibitem{Yu2009nnano}
Y-J. Yu, Y. Zhao, S. Ryu, L. E. Brus, K. S. Kim and P. Kim,
\emph{Nano Lett.}  \textbf{9}, 3430  (2009).

\bibitem{Bocquillon2013science}
E. Bocquillon, V. Freulon,  J-M. Berroir,  P. Degiovanni, B. Pla\c{c}ais, A. Cavanna, Y. Jin, G. F\`eve,
\emph{Science} \textbf{339}, 6123  (2013).

\bibitem{Dubois2013nature}
J. Dubois, T. Jullien, F. Portier, P. Roche, A. Cavanna, Y. Jin, W. Wegscheider, P. Roulleau, D.C. Glattli,
\emph{Nature} \textbf{502}, 7473  (2013).

 \bibitem{Seneor2012mrs}
P. Seneor, B. Dlubak, M.B. Martin, A. Anane, H. Jaffres, A. Fert,
 \emph{MRS Bulletin} \textbf{37}, 1245 (2012).

\bibitem{Katnelson2006nphys}
M. I. Katnelson,  K. S. Novoselov, A. K. Geim,
\emph{Nat. Phys.} \textbf{2}, 620  (2006),

\bibitem{Lee2015nphys}
 G.-H. Lee,	G.-H. Park, H.-J Lee,
\emph{Nat. Phys.}, doi 10.1038 nphys3460  (2015).

\bibitem{Wilmart2014_2dm}
Q. Wilmart, S. Berrada, D. Torrin, V. Hung Nguyen, G. F\`eve, J-M. Berroir, P. Dollfus and B. Placais,
\emph{2D Materials}  \textbf{1},  011006 (2014).

\bibitem{Huard2007prl}
B. Huard, J. A. Sulpizio, N. Stander, K. Todd, B. Yang,  D. Goldhaber-Gordon,
\emph{Phys. Rev. Lett.} \textbf{98}, 236803 (2007).

\bibitem{Standler2009PRL}
N. Stander, B. Huard, and D. Goldhaber-Gordon,  \emph{Phys. Rev. Lett.} \textbf{102}, 026807 (2009),

\bibitem{Rickhaus2013ncomm}
 P. Rickhaus, R. Maurand, M-H Liu, M. Weiss, K. Richter, C. Sch\"onenberger, \emph{Nature Communications} \textbf{4}, 2342 (2013).

\bibitem{Meric2011ieee}
I. Meric, C.R. Dean, SJ Han, L. Wang, K.A. Jenkins, J. Hone, K.L. Shepard,
\emph{IEEE International Electron Devices Meeting},  978-1-4577-0505-2 (2011).

\bibitem{Gomez2014_2dm}
A. Castellanos-Gomez, M. Buscema, R. Molenaar, V. Singh, L. Janssen, H.S.J. van der Zant and G.A. Steele
\emph{2D Materials}  \textbf{1},  011002 (2014).

\bibitem{Hattori2015acsnano}
Y. Hattori, T. Taniguchi, K. Watanabe, and K. Nagashio,
\emph{ACS Nano.} 9, 916 (2015).

\bibitem{Gorbachev2008nl}
R. V. Gorbachev, A. S. Mayorov, A.S. Savchenko, D.W. Horsell, F. Guinea,
\emph{Nano Lett.} \textbf{8}, 1995 (2008)

\bibitem{Young2009nphys}
A. F. Young, P. Kim, \emph{Nat. Phys.} \textbf{5}, 222  (2009),

\bibitem{Zhong2015nr}
H. Zhong, Z. Zhang, B. Chen, H. Xu, D. Yu, L. Huang, L-M. Peng, \emph{Nano Research } \textbf{8}, 1669 (2015)

\bibitem{Pallecchi2011prb}
E. Pallecchi,J. Chaste, G. F\`eve, B. Huard, T. Kontos, J-M. Berroir,  B. Pla\c{c}ais,  \emph{Phys. Rev. B} \textbf{83}, 125408 (2011).

\bibitem{Cheianov2006prb}
V.V. Cheianov, V. Falko, \emph{Phys. Rev. B} \textbf{74}, 041403(R) (2006).

\bibitem{Sonin2008prb}
E. B. Sonin, \emph{Phys. Rev. B} \textbf{77}, 233408 (2008).

\bibitem{Lin2010Science}
Y.-M. Lin, C. Dimitrakopoulos, K. A. Jenkins, D. B. Farmer, H.-Y. Chiu, A. Grill, Ph. Avouris,
\emph{Science} \textbf{327}, 662 (2010).

\bibitem{Pallecchi2011apl}
E. Pallecchi, C. Benz, A.C. Betz, H.v. L\"ohneysen, B. Pla\c{c}ais,  R. Danneau, \emph{Appl. Phys. Lett.} \textbf{99}, 113502 (2011).


\end{thebibliography}
\end{document}